\documentclass[aps,pra,twocolumn,showpacs,preprintnumbers]{revtex4-1}

\usepackage{psfrag,graphicx}
\usepackage{dcolumn}
\usepackage{amsmath,amssymb}
\usepackage{amsfonts}
\usepackage{bm}
\usepackage{amsfonts,amssymb,amsmath}
\usepackage{bbold}
\usepackage{epstopdf}
\usepackage{color}
\usepackage{centernot}

\usepackage{verbatim}

\newcommand{\be}{\begin{equation}}
\newcommand{\ee}{\end{equation}}
\newcommand{\bq}{\begin{eqnarray}}
\newcommand{\eq}{\end{eqnarray}}

\DeclareSymbolFont{bbold}{U}{bbold}{m}{n}
\DeclareSymbolFontAlphabet{\mathbbold}{bbold}

\usepackage[caption=false]{subfig}

\begin{document}
\title{Universal quantum computation in the surface code using non-Abelian islands}
\author{Katharina Laubscher, Daniel Loss, and James R. Wootton}
\affiliation{Department of Physics, University of Basel, Klingelbergstrasse 82, CH-4056 Basel, Switzerland}

\begin{abstract}
	
The surface code is currently the primary proposed method for performing quantum error correction. However, despite its many advantages, it has no native method to fault-tolerantly apply non-Clifford gates. Additional techniques are therefore required to achieve universal quantum computation. Here we propose a hybrid scheme which uses small islands of a qudit variant of the surface code to enhance the computational power of the standard surface code. This allows the non-trivial action of the non-Abelian anyons in the former to process information stored in the latter. Specifically, we show that a non-stabilizer state can be prepared, which allows universality to be achieved.
	
\end{abstract}

\maketitle

\section{Introduction}
The excellent properties of surface codes make them among the prime examples of quantum error correcting codes~\cite{Dennis2002}. These codes are the focus of many designs for large scale quantum computation~\cite{Fowler2012,Wootton2012}, as well as the experimental development of fault-tolerance~\cite{Kelly2014,Corcoles2015,Takita2016}. Unfortunately, surface codes do have a major disadvantage. Though all multi-qubit Clifford operations can be applied fault-tolerantly using a combination of transversal gates and code deformation~\cite{Brown2017}, there is no similar means to implement a non-Clifford gate. The standard way to achieve universality is then by magic state distillation~\cite{Bravyi2005}. Though work in this direction is very encouraging, it is nevertheless important to maintain discovery of alternatives. Since the exact constraints of the devices that will achieve fault-tolerance are not yet fully known, we must make sure to be prepared to tailor to their needs by having a range of directions to pursue. In this work, we therefore present an alternative to magic state distillation which builds upon the fact that the disadvantage described above only strictly applies to surface codes of \emph{qubits}.

Generalized codes can be defined using a lattice of $|G|$-level qudits, where $G$ is a finite group. The stabilizer operations are now based on the algebraic structure of $G$. These codes are known as \emph{quantum double models}~\cite{Kitaev2003}.  In particular, when $G$ is a non-Abelian group, the code plays host to non-Abelian anyons.

Standard surface codes are based on the group $\mathbb{Z}_2$. They are realized using qubits, which are the most well-developed quantum system for quantum computation. The codes also require only nearest-neighbour operations on a two-dimensional lattice: a realistic layout that is already being tested in some prototype devices~\cite{IBM,google}. Furthermore, the problem of decoding a syndrome for these codes corresponds to minimum weight perfect matching, for which highly fast and effective algorithms are known~\cite{Dennis2002}. When any more complex group is used, some or all of these advantages will be lost: qudits must be used, strict nearest-neighbour interactions might be sacrificed in order to factorize into simpler qudits, and the decoding problem is more complex~\cite{Wootton2008,Hutter2015,Wootton2016}. However, a major advantage is gained: codes based on non-Abelian groups allow for a universal set of gates to be implemented~\cite{Mochon2004}. In order to benefit from the advantages of surface codes, but also obtain the universality of non-Abelian codes, we propose a hybrid approach: A surface code is used to store and manipulate logical qubits, with small islands of non-Abelian codes used as factories producing non-stabilizer states. Specifically, we use the quantum double model based on $S_3$, the permutation group of three objects. $S_3$ is the smallest non-Abelian group and therefore yields the simplest realization of a non-Abelian quantum double model. The required six-level systems can be realized by suitably entangling qubits with qutrits. In particular, one could envision to realize the qutrits by $\mathbb{Z}_3$ parafermions~\cite{Hutter2016}.

\section{The surface code}
The standard $D(\mathbb{Z}_2)$ surface code is defined on a planar square lattice with a single qubit placed on each edge. For each vertex $v$ and each plaquette $p$, we define a stabilizer operator
\begin{align}
A_v^{\mathbb{Z}_2}&=\frac{1}{2}\Bigg(\mathbb{1}+\prod_{j\in\mathrm{star}(v)} \sigma_j^x\Bigg),\label{eq:vertexZ2}\\
B_p^{\mathbb{Z}_2}&=\frac{1}{2}\Bigg(\mathbb{1}+\prod_{j\in\partial p}\ \sigma_j^z\Bigg),\label{eq:plaquetteZ2}
\end{align}
where $\mathrm{star}(v)$ denotes the qubits adjacent to $v$, $\partial p$ denotes the boundary of $p$ and $\sigma^a_j$ for $a\in\{x,y,z\}$ denotes the usual single qubit Pauli-$X$, Pauli-$Y$ and Pauli-$Z$ operators acting on qubit $j$. The Hamiltonian is then defined as $H^{\mathbb{Z}_2}=-\sum_v A_v^{\mathbb{Z}_2} -\sum_p\  B_p^{\mathbb{Z}_2}$, where the sums run over all vertices and plaquettes of the lattice, respectively. The ground state of $H$ is given by the simultaneous $+1$ eigenspace of all stabilizer operators, and is non-degenerate on an infinite lattice without any boundaries.
Excitations corresponding to local violations of stabilizers can be interpreted as anyons due to their non-trivial braiding statistics~\cite{Kitaev2003}. In particular, the $D(\mathbb{Z}_2)$ anyon model consists of four anyons $\{1,e,m,\epsilon\}$. Here, $1$ denotes the vacuum, while $e$ ($m$) anyons live on vertices (plaquettes) of the lattice and correspond to a $+1$ eigenvalue of a projector%
\begin{align}
P_v^e&=\frac{1}{2}\Bigg(\mathbb{1}-\prod_{j\in\mathrm{star}(v)} \sigma_j^x\Bigg),\\
P_p^m&=\frac{1}{2}\Bigg(\mathbb{1}-\prod_{j\in\partial p}\ \sigma_j^z\Bigg),
\end{align}
respectively. The $\epsilon$ particle corresponds to the composite $e\times m$. The $D(\mathbb{Z}_2)$ excitations can be moved by acting on the lattice qubits with strings of Pauli operators~\cite{Kitaev2003}, such that fusion as well as braiding of quasiparticles can be performed in the surface code architecture. The corresponding braiding and fusion rules can easily be found in the literature and will not be summarized here.

While the surface code is conveniently described in terms of the Hamiltonian $H^{\mathbb{Z}_2}$, it is challenging to realize the required four-body interactions in practice. A more realistic approach uses periodically repeated stabilizer measurements to simulate the corresponding system~\cite{Dennis2002}. These measurements detect errors manifesting in the form of unwanted quasiparticles, which can then be removed by suitable decoding algorithms. A summary of error detection and correction procedures can be found in~\cite{Nielsen2000,Lidar2013}.

\begin{figure}[tb]
	\centering
	\includegraphics[scale=1.2]{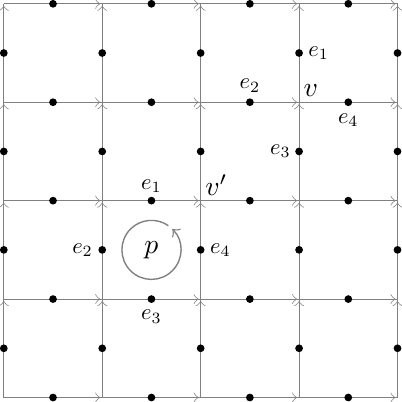}
	\caption{The $D(S_3)$ quantum double model is defined on an oriented square lattice. Six-level qudits, shown as dots, are situated at each edge. For each vertex and each plaquette, we define operators $A_{v}^g$ and $B_{p,v'}^h$ acting on the lattice qudits according to Eqs.~(\ref{eq:vertexS3}) and (\ref{eq:plaquetteS3}).}
	\label{fig:lattice}
\end{figure}

\section{The $D(S_3)$ quantum double model}
Let us now turn to the simplest non-Abelian quantum double model $D(S_3)$, where $S_3$ denotes the permutation group of three objects. We write $S_3=\{e,t,c,ct,c^2,c^2t\}$, where $e$ denotes the identity element and the two generators $t$ and $c$ satisfy $ct=tc^2$, $t^2=c^3=e$. The $D(S_3)$ quantum double model is then defined on a two-dimensional oriented square lattice. On each edge resides a physical six-level system which takes values in the group algebra $\mathbb{C}[S_3]$ with orthonormal basis $(|g\rangle|\,g\in S_3)$. We envision that such a system may, e.g., be realized by suitably entangling a qubit and a $\mathbb{Z}_3$ parafermion~\cite{Hutter2016}, see also Appendix C. For simplicity, let us fix the lattice orientation together with an enumeration convention for the edges as shown in Fig.~\ref{fig:lattice}. Vertex and plaquette operators are then defined as
\begin{align}%
&A_v^g\ \ =R^{g^{-1}}(e_1)L^{g}(e_2)L^{g}(e_3)R^{g^{-1}}(e_4),
\label{eq:vertexoperator}\\
&\begin{aligned}
B_{p,v}^h&=\sum_{h_1h_2h_3h_4=h}|h_1\rangle_{e_1}\langle h_1||h_2\rangle_{e_2}\langle h_2|\\ &\qquad\qquad\qquad\times|h_3^{-1}\rangle_{e_3}\langle\ h_3^{-1}||h_4^{-1}\rangle_{e_4}\langle h_4^{-1}|\label{eq:plaquetteoperator}
\end{aligned}
\end{align}
for $g, h\in S_3$. Here, $R^g$ ($L^g$) denotes the right (left) group multiplication acting on qudit $j$ located at the edge $e_j$. $A_v^g$ depends on the orientation of the lattice in the following way: If $e_j$ points towards (away from) $v$, we choose $L^{g}$ ($R^{g^{-1}}$) to act on qudit $j$. The product in $B_{p,v}^h$ is taken in counterclockwise order starting from and ending at $v$. Again, the orientation of the lattice is taken into account by projecting the state of qudit $j$ onto $h$ ($h^{-1})$ if $e_j$ points in clockwise (counterclockwise) direction when surrounding $p$. We now introduce a set of mutually commuting projectors%
\begin{align}
A_v^{S_3}&=\frac{1}{6}\sum_{g\in S_3} A_v^g,\label{eq:vertexS3}\\
B_p^{S_3}&=B^e_{p,v},\label{eq:plaquetteS3}
\end{align}
which we will refer to as the $D(S_3)$ stabilizer operators. The total Hamiltonian is then defined as
$H^{S_3}=-\sum_v A_v^{S_3}-\sum_p B_p^{S_3}.$

Again, local violations of stabilizer conditions can be interpreted as quasiparticles. In total, the $D(S_3)$ anyon model consists of 8 particles, labelled $A$ through $H$, which live on vertices, plaquettes, or combinations thereof~\cite{Kitaev2003}. A short review of the $D(S_3)$ quantum double model is given in Appendices A and B. For our current purpose, however, it is sufficient to work with the closed submodel $\{A, B, G\}$. Here, $A$ denotes the vacuum, $B$ is an Abelian anyon and $G$ is a non-Abelian anyon reflecting the fact that $S_3$ is a non-Abelian group. In the lattice architecture, these anyons correspond to a $+1$ eigenvalue of the projectors
\begin{align}
&P_v^A\ =\frac{1}{6}\big(A_v^e+A_v^t+A_v^c+A_v^{ct}+A^{c^2}_v+A^{c^2t}_v\big),
\\
&P_v^B\ =\frac{1}{6}\big(A_v^e-A_v^t+A_v^c-A_v^{ct}+A^{c^2}_v-A^{c^2t}_v\big),\label{eq:Bprojector}
\\&\begin{aligned}
P^G_{p,v}&=\frac{1}{3}\big[B^c_{p,v}\big(A^e_v+\omega A^c_v+\bar{\omega} A^{c^2}_v\big)\\ &\qquad+B^{c^2}_{p,v}\big(A^e_v+\bar{\omega} A^c_v+\omega A^{c^2}_v\big)\big],\label{eq:Gprojector}
\end{aligned}
\end{align}
where $\omega=e^{\frac{2\pi i}{3}}$. They can be created, moved and fused following the protocols presented in Refs.~\cite{Brennen2009,Luo2011}. The non-trivial fusion rules are given by $B\times B=A$, $B\times G=G$ and $G\times G=A+B+G$. 
For brevity, we will not give the full set of braiding matrices and instead refer the interested reader to Ref.~\cite{Cui2015}. As in the $D(\mathbb{Z}_2)$ case, the $D(S_3)$ anyons can be detected via syndrome measurements instead of using a Hamiltonian, see Appendix C.

\section{A domain wall between $D(\mathbb{Z}_2)$ and $D(S_3)$}
\begin{figure}[tb]
		\centering
		\includegraphics[scale=1.2]{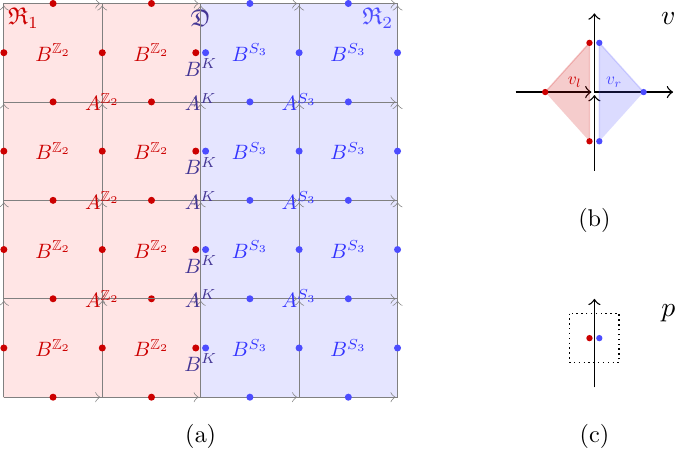}%
		\caption{(a) The domain wall $\mathfrak{D}$ separates the lattice into two regions. Region $\mathfrak{R}_1$ ($\mathfrak{R}_2$) is associated with the $D(\mathbb{Z}_2)$ [$D(S_3)$] quantum double model. Qubits (six-level qudits) are indicated as red (blue) dots. The corresponding stabilizer operators act on the vertices and plaquettes as indicated. (b) A domain wall vertex consists of two `halves' $v_l$ and $v_r$, where $v_l$ ($v_r$) belongs to $\mathfrak{R}_1$ ($\mathfrak{R}_2$). (c) A domain wall plaquette consists of a single qubit and a single six-level spin.}%
		\label{fig:domainwall_stabilizers}%
	\end{figure}

Let us now consider a domain wall separating the lattice into two regions $\mathfrak{R}_1$ and $\mathfrak{R}_2$, where $\mathfrak{R}_1$ ($\mathfrak{R}_2$) is associated with the $D(\mathbb{Z}_2)$ [$D(S_3)$] quantum double model. The domain wall is chosen to lie on the edges of the lattice, and both a qubit and a six-level spin are placed on each domain wall edge, see Fig.~\ref{fig:domainwall_stabilizers}(a). The possible gapped domain walls for this setup are parametrized by the subgroups $K$ of $\mathbb{Z}_2\times S_3$ and a 2-cocycle $\varphi\in H^2(K,\mathbb{C}^\times)$~\cite{Beigi2011}. For reasons that will become clear later, we choose the subgroup $K=\{(e,e),(e,c),(e,c^2),(x,t), (x,ct), (x,c^2t)\}$ with the trivial 2-cocycle and where we used the notation $\mathbb{Z}_2=\{e,x\}$. The domain wall vertex operators $A_v^K$ are then defined as
\begin{equation}
\label{eq:boundaryvertexprojectorZ2S3}
\begin{split}
A^K_v&=\frac{1}{6} (A^e_{v_l}\otimes A^e_{v_r}+A^e_{v_l}\otimes A^c_{v_r}+A^e_{v_l}\otimes A^{c^2}_{v_r}\\&\qquad+A^x_{v_l}\otimes A^t_{v_r}+A^x_{v_l}\otimes A^{ct}_{v_r}+A^x_{v_l}\otimes A^{c^2t}_{v_r}),
\end{split}
\end{equation}
where the notations $v_l$ ($v_r$) for the left (right) half of a vertex are visualized in Fig.~\ref{fig:domainwall_stabilizers}(b). 
The vertex operators for the $D(\mathbb{Z}_2)$ part of the lattice are given by $A_v^e=\mathbb{1}$ and $A_v^x=\prod_{j\in\mathrm{star}(v)}\sigma_j^x$, whereas the vertex operators for the $D(S_3)$ part are defined analogously to Eq.~(\ref{eq:vertexoperator}), but acting on three qudits instead of four. The domain wall plaquette operators $B^K_p$ act on a single qubit and a single six-level spin, see Fig.~\ref{fig:domainwall_stabilizers}(c), and explicitly read
\begin{equation}
\label{eq:boundaryplaquetteprojectorZ2S3}
\begin{split}
B^K_p&=|e,e\rangle\langle e,e|+|e,c\rangle\langle e,c|+|e,c^2\rangle\langle e,c^2|\\&\quad+|x,t\rangle\langle x,t|+|x,ct\rangle\langle x,ct|+|x,c^2t\rangle\langle x,c^2t|.
\end{split}
\end{equation}
%
The domain wall stabilizers commute with each other as well as with neighbouring $D(\mathbb{Z}_2)$ and $D(S_3)$ stabilizers. We now define the total Hamiltonian as $H=H^{S_3}(\mathfrak{R}_1)+H^{\mathbb{Z}_2}(\mathfrak{R}_2)+H^K(\mathfrak{D}),$
where $H^K=-\sum A^K_v-\sum B^K_p$. Figure~\ref{fig:domainwall_stabilizers}(a) graphically summarizes these different terms. Suitable measurement circuits for the domain wall stabilizers as well as a possible simplification of the model in the absence of domain wall plaquette excitations can be found in Appendix C.

The anyon models in $\mathfrak{R}_1$ and $\mathfrak{R}_2$ are related by a set of condensation rules determining how the different species of quasiparticles translate into each other upon crossing the domain wall. For a given domain wall, these rules can be obtained using general algebraic arguments~\cite{Beigi2011,Cong2016}. The condensations which will be relevant for our considerations, however, can easily be understood even without rigorous calculations. Let us focus on the case of a quasiparticle crossing the domain wall from $\mathfrak{R_2}$ to $\mathfrak{R_1}$. Trivially, it is clear that $A$ condenses to $1$. Let us now consider a $B$ anyon. Equation~(\ref{eq:Bprojector}) implies that this anyon is moved by a string of single qudit operations of the form $\sum_{g\in\{e,c,c^2\}}|g\rangle\langle g|-\sum_{g\in\{t,ct,c^2t\}}|g\rangle\langle g|$. Once the $B$ anyon has reached a domain wall vertex $v_d$, the only operation that can move it away from $v_d$ and into $\mathfrak{R_1}$ is a Pauli-$Z$ operation on a qubit in $\mathfrak{R_1}$. This operation, however, creates an $e$ particle on a neighbouring vertex. Thus, we arrive at the conclusion that $B$ condenses to $e$. Using similar considerations, we find that the $G$ anyon cannot enter the $D(\mathbb{Z}_2)$  phase. A generalized set of condensation rules, where the full $D(S_3)$ model is taken into account, can be found in Appendix D.

\section{Processing information with Abelian $D(\mathbb{Z}_2)$ anyons}
The standard technique to encode information in the surface code involves the creation of holes~\cite{Raussendorf2007,Fowler2009,Fowler2012}, where a number of either plaquette or vertex stabilizers are removed from a given region of the lattice. Since these stabilizers are absent from the Hamiltonian (no longer measured by the stabilizer circuits), anyons may be `absorbed' by the hole, i.e., they condense to the vacuum upon entering the hole. In particular, a hole which resides on the vertices (plaquettes) of the lattice may absorb an $e$ ($m$) anyon. We will call these holes $1+e$ ($1+m$) holes, according to the types of anyons that condense at their boundaries. The anyonic occupancy of an $e$ ($m$) hole defines a two-dimensional Hilbert space $\mathrm{span}(|1\rangle,|e\rangle)$ [$\mathrm{span}(|1\rangle,|m\rangle)$], which can be used to encode a qubit. For practical reasons, however, one usually uses a \emph{hole-pair} encoding, where a qubit is encoded in $\mathrm{span}(|11\rangle,|ee\rangle)$ [$\mathrm{span}(|11\rangle,|mm\rangle)$] of two holes with trivial total occupancy. There exist explicit protocols that fault-tolerantly implement the initialization, movement and read-out of hole-encoded qubits~\cite{Fowler2009,Fowler2012}. In particular, holes can be braided with anyonic excitations or with each other in order to process information in a topologically protected way. Together with other fault-tolerant surface code operations, the full Clifford group $\mathcal{C}_n=\{U \mathrm{\ unitary}\,|\,UPU^{-1}\in\mathcal{P}_n\ \ \forall P\in\mathcal{P}_n\},$ where $\mathcal{P}_n=\{\pm 1,\pm i\}\times\{\mathbb{1},\sigma^x,\sigma^y,\sigma^z\}^{\otimes n}$ is the Pauli group on $n$ qubits, can be realized~\cite{Brown2017}.

While the Clifford operations on their own are not sufficient for universal quantum computation, the additional injection of suitable ancillary states can be used to construct a universal gate set~\cite{Bravyi2005}. In particular, the states of interest are those which cannot be prepared by Clifford operations alone. These states are commonly referred to as \emph{non-stabilizer} states. The remainder of this manuscript will demonstrate how a non-stabilizer state can be prepared using an island of $D(S_3)$ code, and how it can be injected into the standard surface code phase.

\section{Processing information with non-Abelian anyons of the $\{A, B, G\}$ submodel}
The possible fusion outcomes of a pair of $G$ anyons span a three-dimensional Hilbert space which is commonly referred to as the \emph{fusion space}. Due to its inherent non-locality, the fusion space of non-Abelian anyons proves as a convenient place to store quantum information. In the following we consider four $G$ anyons at positions 1 through 4 with total fusion outcome vacuum. We can label their collective state by specifying the intermediate fusion outcomes for a fixed fusion sequence. In particular, we will write the state where $G_1$ and $G_2$ fuse to $x\in \{A,B,G\}$ as $|x\rangle$. Note that the fusion outcome of $G_3$ and $G_4$ is now fixed to $x$ as well by the condition on the total fusion outcome. Thus, the total fusion space is again three-dimensional. We now define braiding matrices $B_{ij}$ describing the exchange of the anyons at positions $i$ and $j$. In general, such an exchange acts non-trivially on the fusion space. While the set of operations generated in this way is far from universal, it allows us to prepare a particular non-stabilizer state. For this, consider the double exchange $B_{23}^2$ which is given by

\begin{equation}
B_{23}^2=e^{\frac{-2i\pi}{9}}\begin{pmatrix}
\cos\left(\frac{2\pi}{3}\right)&i\,\sin\left(\frac{2\pi}{3}\right)&0\\
i\,\sin\left(\frac{2\pi}{3}\right)&\cos\left(\frac{2\pi}{3}\right)&0\\
0&0&e^{\frac{2\pi i}{3}}
\end{pmatrix}
\end{equation}
in the basis $(|A\rangle,|B\rangle,|G\rangle)$~\cite{Cui2015}. This operation preserves the two-dimensional subspace $\mathrm{span}(|A\rangle, |B\rangle)$ and its orthogonal complement $\mathrm{span}(|G\rangle)$. Thus, $B_{23}^2$ is a well-defined operation on a qubit encoded in $\mathrm{span}(|A\rangle, |B\rangle)$, and it is easy to check that this operation is not part of the Clifford group. Creating two pairs of $G$ anyons from the vacuum and applying $B_{23}^2$ now allows us to prepare the non-stabilizer state
\begin{equation}
\label{eq:nonstabilizer}
|\psi\rangle=\cos\left(\frac{2\pi}{3}\right)|A\rangle+i\,\sin\left(\frac{2\pi}{3}\right)|B\rangle,
\end{equation}
ignoring a global phase factor.

\begin{figure*}[tb]
	\centering
	\includegraphics[scale=1]{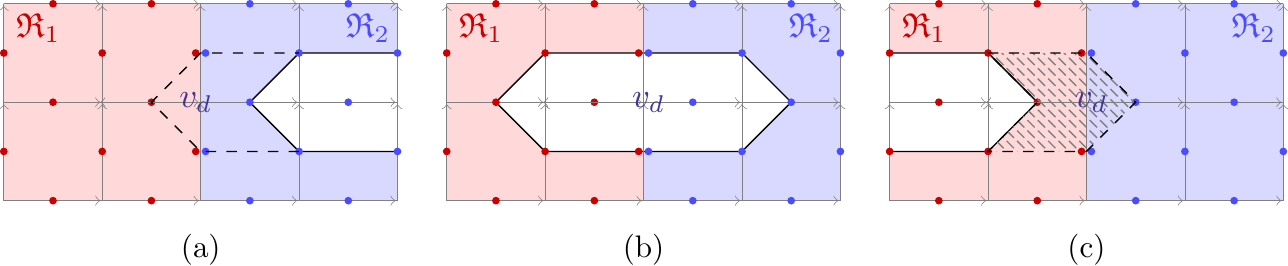}
	\caption{Let $\mathfrak{R_1}$ ($\mathfrak{R_2}$) be associated with the $D(\mathbb{Z}_2)$ [$D(S_3)$] quantum double model. (a) An $A+B+2C$ hole living in $\mathfrak{R_2}$ is moved towards the domain wall. Upon removing the domain wall stabilizer $A_{v_d}^K$, the condensation rules ensure that the logical information is delocalized both in the $D(\mathbb{Z}_2)$ as well as the $D(S_3)$ part of the hole, and that no mixing between basis states occurs. (b) In the intermediate stage of the transfer, the hole has support in both $\mathfrak{R_1}$ and $\mathfrak{R_2}$. (c) The transfer is completed by measuring $A_{v_d}^K$ and applying appropriate error correction procedures, leaving us with a $1+e$ hole living in $\mathfrak{R_1}$.}%
	\label{fig:transfer}
\end{figure*}

\section{Transferring information between the codes}
While the fusion space encoding is most natural for non-Abelian anyon models, a hole encoding similar to the one introduced for the standard surface code is also possible~\cite{Cong2016}. As there exists no fusion space encoding for Abelian models, the state $|\psi\rangle$ given in Eq.~(\ref{eq:nonstabilizer}) has to be converted to such a hole encoding before it can be injected into the $D(\mathbb{Z}_2)$ phase. In particular, we choose to switch to an encoding defined via the anyonic occupancy of two $A+B+2C$ holes. Here, we have adopted the same labelling convention for holes as in the $D(\mathbb{Z}_2)$ case, i.e., we label a hole by the particles that condense at its boundary. The $C$ particle is a non-Abelian anyon living on the vertices of the $D(S_3)$ model, and the factor 2 denotes a condensation multiplicity related to local degrees of freedom associated with this type of particle, see Appendix A.  $A+B+2C$ holes can be created by removing the vertex stabilizers in a given region of the $D(S_3)$ part of the lattice. By fusing each pair of $G$ particles with an empty $A+B+2C$ hole, their anyonic occupancy is transferred to the hole. A more detailed description of this process with particular focus on the topological protection of the encoded information during all stages of the transfer can be found in Appendix E.

Next, we move the $A+B+2C$ holes across the domain wall. Our particular choice of domain wall stabilizers and the emerging condensation rules ensure that an $A+B+2C$ hole in the state $a|A\rangle+b|B\rangle$ will turn into a $1+e$ hole in the state $a|1\rangle+b|e\rangle$ upon entering the $D(\mathbb{Z}_2)$ phase. In particular, suppose that an $A+B+2C$ hole is moved towards the domain wall by expansion and contraction processes~\cite{Cong2016} until the next step in the movement requires the hole to enclose the domain wall vertex $v_d$, see Fig.~\ref{fig:transfer}(a). Crucially, the information stored in the $A+B+2C$ hole must now be delocalized not only in the $D(S_3)$ part of the hole, but also in the part stretching into the $D(\mathbb{Z}_2)$ phase. Additionally, we have to make sure that this process does not corrupt the encoded information by condensation processes that mix the logical basis states. Both of these requirements are met, since the $A$ ($B$) particle enters the $D(\mathbb{Z}_2)$ phase \emph{exclusively} as $1$ ($e$). By extending the hole further into the $D(\mathbb{Z}_2)$ phase, see Fig.~\ref{fig:transfer}(b), and eventually contracting the $D(S_3)$ part, we arrive at the situation shown in Fig.~\ref{fig:transfer}(c). In order to fully contract the $D(S_3)$ part of the hole, the domain wall stabilizer $A_{v_d}^K$ has to be measured. An outcome corresponding to a non-trivial quasiparticle has to be corrected as part of the non-Abelian decoding of the $D(S_3)$ code, but also taking into account possible contributions from the $D(\mathbb{Z}_2)$ part. Appendix F comments on some additional details regarding the fault-tolerance of the injection process. Applying the procedure described here to the state $|\psi\rangle$, we have found a means to produce $1+e$ holes in the non-stabilizer state $|\psi\rangle=\cos\left(\frac{2\pi}{3}\right)|1\rangle+i\,\sin\left(\frac{2\pi}{3}\right)|e\rangle$. These can now be used to perform the non-Clifford operation 
\begin{equation}
U=\begin{pmatrix}
1&0\\
0&e^{\frac{2\pi i}{3}}
\end{pmatrix}
\end{equation}
on the space $\mathrm{span}(|1\rangle,|e\rangle)$ following the arguments of Ref.~\cite{Bravyi2005}, which are summarized in Appendix G for completeness.

\section{Conclusions and Outlook}
While the theory of domain walls between different quantum double models is well established, we have presented an explicit application of such a domain wall for computational purposes. In particular, we have demonstrated the ability to create a non-stabilizer state in the qubit surface code by transferring information with a patch of qudit surface code based on the non-Abelian group $S_3$. Combined with the Clifford gates of the surface code~\cite{Brown2017}, this provides universal quantum computation. 

Fault-tolerance for this scheme depends on the ability to efficiently and effectively decode the non-Abelian syndrome of the $D(S_3)$ quantum double model. Specifically, it must be shown that the logical error rate can be reduced arbitrarily by increasing the size of the islands. While it is widely believed that this can be done for such systems of non-Abelian anyons, most studies have only been done for certain special cases~\cite{Wootton2008,Hutter2015,Wootton2016,Burton2017}. The only current proof for full fault-tolerance concerns so-called \emph{non-cyclic} anyon models~\cite{Dauphinais2017}. The anyons of the $D(S_3)$ code, as well as many other interesting and important anyon models, are not of this type. Our work therefore provides additional motivation for the future study of whether fault-tolerance is possible for \emph{cyclic} anyon models, and how the decoding may be optimally performed.

The six-level qudits required in the non-Abelian part of our model can potentially be realized by entangling a qubit and a $\mathbb{Z}_3$ parafermion. The recent efforts put into the theoretical realization of parafermion zero modes in condensed matter systems, see Ref.~\cite{Alicea2016} for a review, support the hope that models like the one presented here are indeed of significant interest to the future of topological quantum computation.

\acknowledgements
This work was supported by the Swiss National Science Foundation (Switzerland) and the NCCR QSIT.

\appendix

\section{The full $D(S_3)$ quantum double model}

In this Appendix we briefly summarize the theory of the full $D(S_3)$ quantum double model. While the $\{A,B,G\}$ submodel considered in the main text is closed under fusion and braiding and can in principle be seen as an anyonic model on its own, error detection and correction procedures for the presented lattice realization may require knowledge of the full particle spectrum. 

The full $D(S_3)$ anyon model consists of 8 particles~\cite{Kitaev2003}. They can be labelled by a conjugacy class $C$ of $S_3$ and an irreducible representation (irrep) of the normalizer $N_C$ of a representative of $C$. For historical reasons, the former is often referred to as the \emph{magnetic flux} of the particle, whereas the latter is called its \emph{electric charge}. Note that the choice of representative is arbitrary as the corresponding normalizers are isomorphic, which justifies the notation $N_C$. In the spin lattice architecture, particles live on sites $s=(p,v)$ consisting of a plaquette $p$ and a neighbouring vertex $v$. The plaquette operators $B_{p,v}^h$ correspond to projections onto the magnetic flux $h$ at $p$ (measured with respect to $v$). The vertex operators $A_v^g$ form a representation of the group $S_3$ on the Hilbert space carried by the lattice qudits in $\mathrm{star}(v)$, and the electric charge at $v$ corresponds to the sector transforming under a given irrep of $S_3$ (or one of its subgroups). Explicitly, $S_3$ has three conjugacy classes $[e]=\{e\}$, $[t]=\{t,ct,c^2t\}$ and $[c]=\{c,c^2\}$ with $N_{[e]}\sim S_3$, $N_{[t]}\sim \mathbb{Z}_2$ and $N_{[c]}\sim \mathbb{Z}_3$. Let now $s=(p,v)$ be a site. The pure electric charges (with trivial magnetic flux $[e]$) at $v$ are given by the irreps of $S_3$. There are two one-dimensional irreps corresponding to the vacuum $A$ (trivial irrep) and the Abelian anyon $B$ (signed irrep), as well as a two-dimensional irrep corresponding to the non-Abelian anyon $C$. The projectors onto a pure charge at $v$ are then given by
\begin{align}
P_v^A&=\frac{1}{6}\big(A_v^e+A_v^t+A_v^c+A_v^{ct}+A^{c^2}_v+A^{c^2t}_v\big),\label{eq:Aprojection}
\\
P_v^B&=\frac{1}{6}\big(A_v^e-A_v^t+A_v^c-A_v^{ct}+A^{c^2}_v-A^{c^2t}_v\big),\label{eq:Bprojection}
\\
P_v^C&=\frac{1}{3}\big(2A^e_v-A^c_v-A^{c^2}_v\big).
\end{align}
If the flux at $p$ is given by the conjugacy class $[t]$, the charge at $v$ is given by one of the two one-dimensional irreps of $\mathbb{Z}_2$. We find the projections onto the different particle types at $s$ to be
\begin{align}
\begin{split}
P_s^D&=\frac{1}{2}\big[B^t_s\big(A^e_v+A^t_v\big)+B^{ct}_s\big(A^e_v+A^{ct}_v\big)\\&\qquad+B^{c^2t}_s\big(A^e_v+A^{c^2t}_v\big)\big],\end{split}\\
\begin{split}
P_s^E&=\frac{1}{2}\big[B^t_s\big(A^e_v-A^t_v\big)+B^{ct}_s\big(A^e_v-A^{ct}_v\big)\\&\qquad+B^{c^2t}_s\big(A^e_v-A^{c^2t}_v\big)\big].
\end{split}
\end{align}
Finally, if the flux at $p$ is $[c]$, the charge at $v$ is given by one of the three one-dimensional irreps of $\mathbb{Z}_3$. The projectors onto the different particle types at $s$ are then given by
\begin{align}
\begin{split}
P^F_s&=\frac{1}{3}\big[B^c_s\big(A^e_v+A^c_v+A^{c^2}_v\big)\\&\qquad+B^{c^2}_s\big(A^e_v+A^c_v+A^{c^2}_v\big)\big],\end{split}\\\begin{split}
P^G_s&=\frac{1}{3}\big[B^c_s\big(A^e_v+\omega A^c_v+\bar{\omega} A^{c^2}_v\big) \\&\qquad+B^{c^2}_s\big(A^e_v+\bar{\omega} A^c_v+\omega A^{c^2}_v\big)\big],\end{split}\\\begin{split}
P^H_s&=\frac{1}{3}\big[B^c_s\big(A^e_v+\bar{\omega} A^c_v+\omega A^{c^2}_v\big)\\&\qquad+B^{c^2}_s\big(A^e_v+\omega A^c_v+\bar{\omega} A^{c^2}_v\big)\big]\end{split}\label{eq:Hprojection}
\end{align}
with $\omega=e^{\frac{2\pi i}{3}}$. Particles with both non-trivial flux and charge are often referred to as \emph{dyons}.

Note that a particle with magnetic flux corresponding to a conjugacy class $C$ can be seen as carrying a $|C|$-dimensional internal Hilbert space spanned by the different elements of $C$. Similarly, a particle with charge corresponding to a $d$-dimensional representation carries a $d$-dimensional internal subspace. As such, the internal space of a $C$, $F$, $G$ or $H$ particle is two-dimensional, whereas the $D$ and $E$ particles carry a three-dimensional internal subspace. As opposed to the type of a particle, these internal degrees of freedom are not topologically protected and can be changed by local operations.

\section{Ribbon operators for the $\{A,B,G\}$ submodel}

\begin{figure}[tb]
	\centering
	\includegraphics{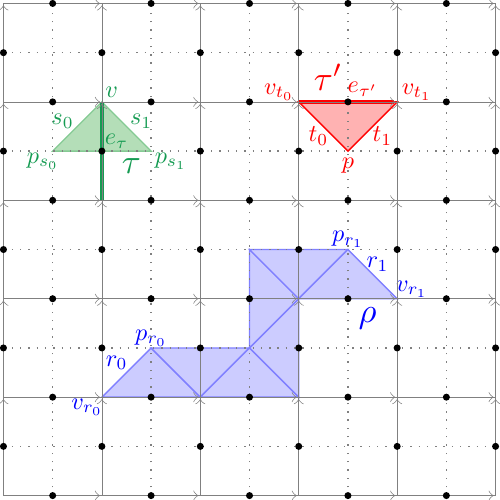}
	\caption{Examples of ribbons. $\tau$ is a dual triangle with endpoints $s_0$ and $s_1$. ‘$\tau'$ is a direct triangle with endpoints $t_0$ and $t_1$. $\rho$ is an arbitrary ribbon composed of 7 triangles and with endpoints given by $r_0$ and $r_1$.}
	\label{fig:ribbons}
\end{figure}

Let us now describe the explicit form of the operators that are required to realize the processes described in the main text. For this, we need to introduce some additional notations. Consider the lattice shown in Fig.~\ref{fig:ribbons}. Adjacent sites can be connected by triangles, where a triangle can belong to one of two types: Two adjacent sites $s_0=(p_{s_0},v)$, $s_1=(p_{s_1},v)$ that share the vertex $v$ are connected by a triangle that is formed by $s_0$, $s_1$ and the dual edge connecting $p_{s_0}$ to $p_{s_1}$. We will call such a triangle a \emph{dual triangle}. Two adjacent sites $t_0=(p,v_{t_0})$, $t_1=(p,v_{t_1})$ that share the plaquette $p$ are connected by a triangle that is formed by $t_0$, $t_1$ and the direct edge connecting $v_{t_0}$ to $v_{t_1}$. We will call such a triangle a \emph{direct triangle}. Every triangle $\tau$ contains exactly one qudit $j_\tau$ situated at a direct edge $e_{\tau}$ of the lattice, see Fig.~\ref{fig:ribbons}. Let us now define operators acting on this single qudit, depending on the type of triangle as well as the orientation of the edge $e_{\tau}$,
\begin{align}
L_{\tau_{\mathrm{dual}}}^h&=\begin{cases} L^{h}(j_\tau) & e_{\tau}\mbox{ points towards }v,\\R^{h^{-1}}(j_\tau) & e_{\tau}\mbox{ points away from }v,\end{cases}\label{eq:directtriangleoperator}\\
P_{\tau_{\mathrm{dir}}}^g&=\begin{cases} |g\rangle_{j_\tau}\langle g| &e_{\tau}\mbox{ points in clockwise} \\&\mbox{direction with respect to } p,\\|g^{-1}\rangle_{j_\tau}\langle g^{-1}| & e_{\tau}\mbox{ points in counterclockwise}\\&\mbox{direction with respect to } p, \end{cases}\label{eq:dualtriangleoperator}
\end{align}
for $h,g\in S_3$. Let us note that there is a certain choice regarding the definition of the starting and ending sites $s_0$ and $s_1$ of a triangle. This can be resolved by always labelling the starting and ending sites of a dual triangle such that $(p_0,p_1,v)$ lists the corners of the triangle in counterclockwise order, and for a direct triangle we will label the starting and ending sites such that $(v_0,v_1,p)$ lists the corners of the triangle in clockwise order. This convention can be reversed for a given triangle by reversing $e_\tau$ each time before the operators (\ref{eq:directtriangleoperator}) or (\ref{eq:dualtriangleoperator}) are applied, and afterwards returning the edge to its original orientation.

In order to connect non-adjacent sites we can combine several triangles to form a \emph{ribbon}. If the ending site $s_1$ of a triangle $\tau$ coincides with the starting site $t_0$ of a second triangle $\tau'$, these triangles can be composed to form the ribbon $\rho=\tau\tau'$. In general, two ribbons are called \emph{composable} if the ending site of the first ribbon coincides with the starting site of the second ribbon. We assume that the dual and the direct part of a ribbon each avoid self-crossing. An example of a ribbon consisting of multiple triangles is also shown in Fig.~\ref{fig:ribbons}.

We now define operators acting on ribbons: For direct or dual triangles, respectively, we let
\begin{align}
F^{h,g}_{\tau_\mathrm{dual}}&=\delta_{g,e}L_{\tau_{\mathrm{dual}}}^h,\\
F^{h,g}_{\tau_\mathrm{dir}}&=P_{\tau_{\mathrm{dir}}}^{g^{-1}}.
\end{align}
The ribbon operators for longer ribbons are then defined recursively via the so-called \emph{glueing relation}
\begin{equation}
\label{eq:glueingrel}
F^{h,g}_{\rho}=\sum_{k\in S_3} F^{h,k}_{\rho_1} F^{k^{-1}hk,k^{-1}g}_{\rho_2},
\end{equation}
where $\rho=\rho_1\rho_2$ for two composable ribbons $\rho_1$, $\rho_2$. In the following we assume that the starting site $s_0$ and the ending site $s_1$ of a ribbon $\rho$ do not overlap. We call such ribbons \emph{open} ribbons. The case of \emph{closed} ribbons, for which we have $s_0=s_1$, will be mentioned later. The most important property of open ribbon operators is that an open ribbon operator $F^{h,g}_{\rho}$ commutes with all terms of the Hamiltonian $H^{S_3}$ except for those corresponding to the starting site $s_0=(p_0,v_0)$ and the ending site $s_1=(p_1,v_1)$ of $\rho$, and it can be shown that all states with excitations only at $s_0$ and $s_1$ can be obtained by applying a linear combination of ribbon operators. Moreover, the action of a ribbon operator $F^{h,g}_{\rho}$ does not explicitly depend on $\rho$, but only on its starting and ending sites $s_0$, $s_1$. In our case, we are interested in the submodel $\{A,B,G\}$. The other particles of the $D(S_3)$ particle spectrum can be treated in a similar way, with the details discussed in Refs.~\cite{Brennen2009,Luo2011}. The $B$ anyon is the simplest non-trivial particle of the $D(S_3)$ particle spectrum as it is the only non-trivial Abelian anyon. Similarly to the particles in the $D(\mathbb{Z}_2)$ quantum double model, the $B$ anyon can be created and moved by applying single qudit unitaries to a path of qudits on the direct lattice. The explicit operator to create a pair of $B$ anyons on two neighbouring vertices $v$, $v'$ is given by a unitary $F^B_{\tau_{\mathrm{dir}}}$ acting on the direct triangle $\tau_{\mathrm{dir}}$ with starting vertex $v$ and ending vertex $v'$,
\begin{equation}
F^B_{\tau_{\mathrm{dir}}}=P_{\tau_\mathrm{dir}}^e-P_{\tau_\mathrm{dir}}^t+P_{\tau_\mathrm{dir}}^c-P_{\tau_\mathrm{dir}}^{ct}+P_{\tau_\mathrm{dir}}^{c^2}-P_{\tau_\mathrm{dir}}^{c^2t}.
\end{equation}
This operator does not depend on the orientation of the corresponding triangle, which is why we can label it by the single spin $j$ that is contained in $\tau_{\mathrm{dir}}$ and write
\begin{equation}
\label{eq:Bribbonop}
\begin{split}
F^B_{j}&=|e\rangle_j\langle e|-|t\rangle_j\langle t|+|c\rangle_j\langle c|\\&\qquad-|ct\rangle_j\langle ct|+|c^2\rangle_j\langle c^2|-|c^2t\rangle_j\langle c^2t|.
\end{split}
\end{equation}
The operator $F^B_j$ can also be employed to move existing $B$ anyons from one vertex to another, which leads to the expression that was given in the main text. In contrast to this, the $G$ particles are non-Abelian dyons, which makes the corresponding creation and movement processes a lot more complicated. In order to create a pair of $G$ anyons on neighbouring, non-overlapping sites we need to apply a ribbon operator $F^{G;(u,u')}_\rho$ for arbitrary local flux degrees of freedom $u,u'\in [c]$ and $\rho=\tau_{\mathrm{dual}}\tau_{\mathrm{dir}}$ for a composable dual and direct triangle. We can, for example, trace out the local degrees of freedom by summing $F^{G;(u,u)}_\rho$ over all elements $u$ of the conjugacy class $[c]$. Explicitly, this corresponds to applying the ribbon operator
\begin{equation}
\label{eq:Gribbonop}
\begin{split}
F^G_{\rho}&=\frac{1}{3}\big[L^c_{\tau_{\mathrm{dual}}}(P^e_{\tau_{\mathrm{dir}}}+\omega P^c_{\tau_{\mathrm{dir}}}+\bar\omega P^{c^2}_{\tau_{\mathrm{dir}}})\\&\qquad+L^{c^2}_{\tau_{\mathrm{dual}}}(P^e_{\tau_{\mathrm{dir}}}+\bar\omega P^c_{\tau_{\mathrm{dir}}}+\omega P^{c^2}_{\tau_{\mathrm{dir}}})\big]
\end{split}
\end{equation}
with $\omega=e^{\frac{2\pi i}{3}}$. The glueing relation then allows us to extend this ribbon to more triangles. Importantly, note that if $\rho$ consists of multiple triangles, $F^G_{\rho}$ cannot be written as a tensor product of operators acting on single triangles. While the operation (\ref{eq:Gribbonop}) is not unitary, it can be constructed using an adaptive procedure as explained in Ref.~\cite{Luo2011}. The same reference also describes how to extend arbitrary ribbon operators in order to move the corresponding particles around the lattice. How the required operations, in particular the ones involving projections, can be applied in practice depends on the particular experimental realization.

Let now $\sigma$ be a closed ribbon, i.e. $s_0=s_1$. It can be shown that there exists a basis for the algebra of ribbon operators on $\sigma$ which corresponds to projections onto a definite topological charge within the region enclosed by $\sigma$~\cite{Kitaev2003,Bombin2008}. We can use this to construct logical operators for the encoding schemes presented in the main text. Explicitly, the projection onto the trivial total charge $A$ within the enclosed region is given by
\begin{equation}
\label{eq:Bribbonprojector}
K^{A}_{\sigma}=\frac{1}{6}( F_{\sigma}^{e,e}+ F_{\sigma}^{t,e}+F_{\sigma}^{c,e}+ F_{\sigma}^{ct,e}+F_{\sigma}^{c^2,e}+ F_{\sigma}^{{c^2}t,e}),
\end{equation}
while the projection onto the total charge $B$ within the enclosed region is given by
\begin{equation}
\label{eq:Bribbonprojector}
K^{B}_{\sigma}=\frac{1}{6}( F_{\sigma}^{e,e}-F_{\sigma}^{t,e}+F_{\sigma}^{c,e}- F_{\sigma}^{ct,e}+F_{\sigma}^{c^2,e}- F_{\sigma}^{{c^2}t,e}).
\end{equation}
These operators are the direct generalization of the projection operators (\ref{eq:Aprojection}) and (\ref{eq:Bprojection}) to more than one vertex. The logical $Z$ operation for the fusion space encoding (the hole encoding) used in the main text is then given by the ribbon operator
\begin{equation}
\label{eq:logZ}
Z_L=K_\sigma^A-K_\sigma^B,
\end{equation}
where $\sigma$ is a closed ribbon enclosing two $G$ particles of the same pair (one of the holes) completely, but no other excitations. On the other hand, the logical $X$ operation for the fusion space encoding (the hole encoding) is given by a ribbon operator $F_{\rho}^B$, where $\rho$ is an open ribbon connecting two $G$ particles of different pairs (the two holes).

\section{Generalized stabilizer circuits}

In order to physically realize the $D(\mathbb{Z}_2)$ quantum double model, Ref.~\cite{Dennis2002} proposed an implementation that simulates the Hamiltonian $H^{\mathbb{Z}_2}$ using local stabilizer circuits. This was further investigated in Refs.~\cite{Fowler2009,Fowler2012}. A brief outline of how these techniques can be generalized to arbitrary $D(G)$ quantum double models was recently given in Ref.~\cite{Cong2016}. Here we adapt a similar approach to the case of $D(S_3)$, additionally commenting upon some issues arising for non-Abelian quantum double models.\par

\begin{figure}[tb]
	\includegraphics[scale=1.]{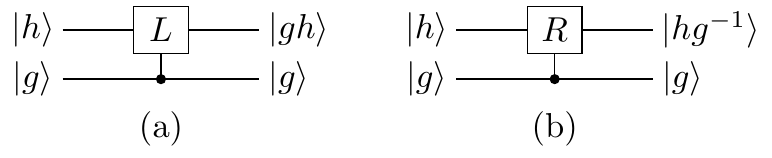}
	\caption{Definition of controlled multiplications for qudits taking values in the group algebra $\mathbb{C}[S_3]$. (a) Definition of the controlled left multiplication. (b) Definition of the controlled right multiplication. Note that by a slight abuse of notation, the gate $R$ is defined as applying the operator $R^{g^{-1}}$ (and not $R^g$) for a control qubit in the state $|g\rangle$.}
	\label{fig:controlledops}
\end{figure}

Let $s=(p,v)$ be a site. According to the definition of the plaquette operators $B_s^h$, see Eq.~(\ref{eq:plaquetteS3}), the flux at $p$ is calculated by taking the ordered product of group elements (or their inverses) corresponding to the states of the qudits in $\partial p$. If the flux at $p$ is $h$, the charge residing at $v$ is given by the irreducible representation according to which the four qudits in $\mathrm{star}(v)$ transform under operations $A_v^g$ for $g\in N_{h}$. This is captured by the projection operators onto the different quasiparticle types at $s$ that were given in Eqs.~(\ref{eq:Aprojection}) to (\ref{eq:Hprojection}). In order to realize suitable stabilizer circuits, let us introduce controlled left and right multiplication gates for arbitrary group elements $g\in S_3$ as shown in Fig.~\ref{fig:controlledops}. These circuits can be considered generalizations of the well-known $\mathrm{CNOT}$ gate, which can be recovered from the definition in Fig.~\ref{fig:controlledops} by replacing $S_3$ by $\mathbb{Z}_2=\{e,x\}$ and setting $g=x$.

\begin{figure}[htb]
	\centering
	\includegraphics{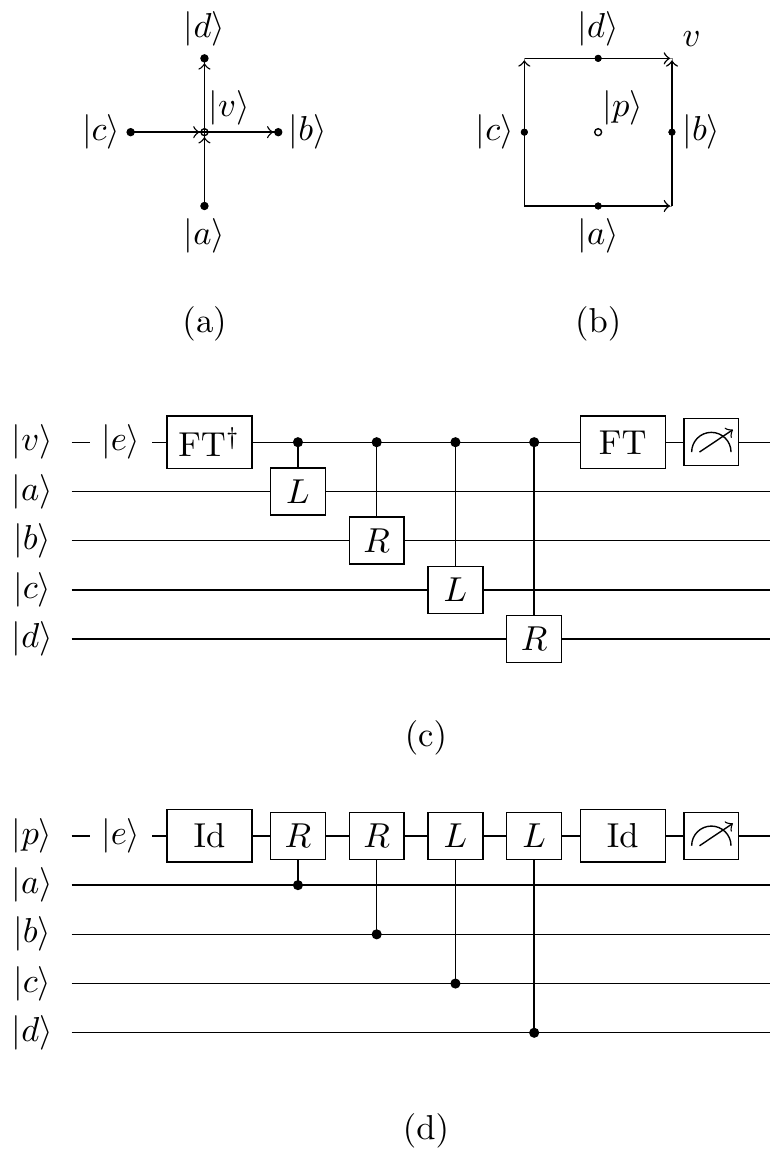}
	\caption{(a) The convention for labelling the data qudits around vertices. The vertex syndrome qudit is denoted as $v$ by a slight abuse of notation, and the data qudits in $\mathrm{star}(v)$ are labelled $a$ to $d$. The zig-zag sequence is crucial in order for neighbouring circuits not to interfere with each other. (b) The convention for labelling the data qudits around plaquettes. The plaquette syndrome qudit is denoted as $p$. Again particular attention must be paid to the specific ordering of the labels. (c) The stabilizer circuit measuring the irreducible representations of $S_3$ according to which the qudits in $\mathrm{star}(v)$ transform under the vertex operations $A_v^g$. The vertex syndrome qudit is denoted as $v$, and the data qudits in $\mathrm{star}(v)$ are labelled $a$ to $d$ in the order indicated in (a). Information on Fourier transforms on non-Abelian groups can be found in Ref.~\cite{Moore2003}. (d) The stabilizer circuit measuring the flux at the plaquette $p$ with respect to the top right neighbouring vertex $v$. The plaquette syndrome qudit is denoted as $p$ and the data qudits in $\partial p$ are labelled $a$ to $d$ in the order indicated in (b).}
	\label{fig:circuits}
\end{figure}

In order to facilitate measuring the quasiparticle occupancy of all lattice sites, let us place an additional ancillary qudit at each plaquette and each vertex. In order to distinguish between the ancillary qudits and the actual qudits comprising the code we call the former \emph{syndrome qudits} and the latter \emph{data qudits}.
We now proceed to construct circuits that simulate the Hamiltonian $H^{S_3}$. In contrast to Abelian models, the projectors $B^h_s$ do not necessarily commute with all the vertex operators $A_v^g$. As a consequence, the projections onto the different particle types cannot, in general, be written as a tensor product of a charge part and a flux part. Complete extraction of the syndrome information is thus not possible via simultaneous flux and charge measurements. However, applying the true quasiparticle projectors would introduce a severe time overhead in the error correction framework. We now argue that if we accept certain limitations concerning the syndrome information that can be obtained, it is still possible to use simultaneous vertex and plaquette measurements. The corresponding circuits are shown in Fig.~\ref{fig:circuits}. Note that the definition of the plaquette circuit shown in Fig.~\ref{fig:circuits}(d) implicitly fixes a choice of sites for the quasiparticles to reside on; by choosing to evaluate all fluxes with respect to the vertex $v$ at the top right corner of $p$, the syndrome information is given in terms of the anyonic occupancy of sites of this form. As in the Abelian case, all circuits operate in lockstep.

Let $s=(p,v)$ denote a particular site of the form specified above. If the order of the controlled operations on the qubits in $\mathrm{star}(v)$ and $\partial p$ is chosen as indicated in Figs.~\ref{fig:circuits}(a) and \ref{fig:circuits}(b), respectively, the circuit for $v$ touches the shared qudits before the circuit for $p$ does. The vertex circuits were designed to measure the irreducible representation of $S_3$ according to which the qudits in $\mathrm{star}(v)$ transform under the vertex operations $A_v^g$. However, for a flux in the conjugacy class $C$ we need to project onto an irreducible representation of $N_C$. Since the restriction of an irreducible representation of $S_3$ onto $N_C$ is not necessarily irreducible, it will in general not be possible to distinguish all the dyons in the particle spectrum of the model. Additionally, operators containing off-diagonal elements of higher-dimensional representations may cause rotations in the local charge subspace carried by a given particle. 

After the vertex circuit, the plaquette circuit touches the shared qudits. The fact that this circuit does not commute with the vertex circuit is reflected in the following way. The local flux eigenstate $|h\rangle$ resulting from the measurement of the plaquette syndrome qudit does indeed correspond to the flux of the particle \emph{after} the full stabilizer circuit cycle. However, this state does not necessarily reflect the original flux state \emph{before} this measurement round, since the vertex circuit at $v$ may have mixed the local flux states of the particle within the conjugacy class $[h]$. Still, the flux measurement gives the correct conjugacy class of the particle located at $(p,v)$ and the correct local flux state after the vertex measurement. Moreover, the local flux state before the vertex measurement can in principle be inferred from the outcome of this measurement and the structure of the irreducible representations. Since we do not use the local degrees of freedom for quantum computational purposes, the rotations that appear as a side effect of the stabilizer circuits do not cause any problems from a theoretical point of view.

Explicitly, it turns out that all the particles of the $D(S_3)$ particle spectrum with flux $[e]$ or $[c]$ can be distinguished by simultaneous plaquette and vertex measurements as defined in Fig.~\ref{fig:circuits}, given that we are willing to accept certain rotations in the local subspaces of the corresponding particles. Indeed, the operations resulting from the vertex measurement circuit defined in Fig.~\ref{fig:circuits}(c) correspond to projections onto irreducible representations of $\mathbb{Z}_3$ combined with rotations in the local charge subspace. The particles $D$ and $E$ corresponding to the conjugacy class $[t]$ cannot always be distinguished in this way.

Additionally, let us briefly comment on the creation and manipulation of holes in the stabilizer circuit framework discussed so far. In the main text we use charge holes residing on the vertices of the lattice in order to store and process information. These are particularly simple to implement. In both the $D(\mathbb{Z}_2)$ as well as the $D(S_3)$ phase they can be created (and extended) by measuring the spins within the boundary of the hole in the computational basis and rotating to the state corresponding to the identity element. The vertex stabilizer circuits inside and on the boundary of the hole are now turned off completely, whereas the plaquette stabilizer circuits operate in the same way as before. Holes can be contracted by turning the vertex stabilizers back on and correcting for non-trivial measurement outcomes. By subsequent expansion and contraction processes, holes can be moved around the lattice.

To conclude this section, let us add a few practical comments. Realizing the $D(S_3)$ model on a lattice requires $6$-level systems that take values in the group algebra $\mathbb{C}[S_3]$, a structure which is unlikely to occur naturally in a physical system. However, the fact that $S_3$ is the semi-direct product of $\mathbb{Z}_3$ and $\mathbb{Z}_2$,
\begin{equation}
S_3=\mathbb{Z}_3\rtimes\mathbb{Z}_2,
\end{equation}
implies that the group structure of $S_3$ can be realized by properly entangling a qutrit and a qubit, which may greatly simplify the experimental realization of the $D(S_3)$ quantum double model. One could, for example, think about using a $\mathbb{Z}_3$ parafermion in order to realize the qutrit. It has been shown that $\mathbb{Z}_3$ parafermions are capable of realizing the full qutrit Clifford group~\cite{Hutter2016}. Explicitly, every element $g\in S_3$ can be written in the form $g=c^rt^s$ with $r\in\{0,1,2\}$ and $s\in\{0,1\}$. The left and right multiplication operators acting on the composite six-level system in terms of the qubit-qutrit basis $|r\rangle|s\rangle=|c^r t^s\rangle$ are then given by
\begin{equation}
\begin{aligned}
&L^e\ \ =\ \mathbb{1}_3\otimes\mathbb{1}_2,\\
&L^t\ \ =\ \mathrm{flip}(1,2)\otimes\sigma_x,\\
&L^c\ \ =\ X\otimes \mathbb{1}_2,\\
&L^{ct}\ =\ \mathrm{flip}(0,1)\otimes\sigma_x,\\
&L^{c^2}\ =\ X^{-1}\otimes \mathbb{1}_2,\\
&L^{c^2t}=\ \mathrm{flip}(0,2)\otimes\sigma_x,\\
&R^e\ \ =\ \mathbb{1}_3\otimes\mathbb{1}_2,\\
&R^t\ \ =\ \mathbb{1}_3\otimes\sigma_x,\\
&R^c\ \ =\ X\otimes |0\rangle\langle 0|+X^{-1}\otimes |1\rangle\langle 1|,\\
&R^{ct}\ =\ X^{-1}\otimes|0\rangle\langle 1|+X\otimes |1\rangle\langle 0|,\\
&R^{c^2}\ =\ X^{-1}\otimes |0\rangle\langle 0|+ X\otimes |1\rangle\langle 1|,\\
&R^{c^2t}=\ X\otimes|0\rangle\langle 1|+X^{-1}\otimes |1\rangle\langle 0|.
\end{aligned}
\end{equation}
Here, $X$ denotes the generalized Pauli-$X$ on a qutrit and $\mathrm{flip}(i,j)$ denotes the operation that swaps the $i$-th and the $j$-th basis state of the qutrit. The single qudit projection operators in the qubit-qutrit basis are simply given by the tensor product of the projections onto the corresponding states of the qubit and the qutrit. These operations can now be used to build up suitable stabilizer circuits.

\section{Gapped domain walls between the $D(S_3)$ phase and the $D(\mathbb{Z}_2)$ phase}
\label{subsec:boundaryz2s3}

In this Appendix, we comment on the considerations that allowed us to find a suitable domain wall for the hybrid architecture introduced in Sec.~IV. Let us again consider a lattice which is separated into two half-planes, where the left (right) half-plane is associated with the $D(\mathbb{Z}_2)$ [$D(S_3)$] quantum double model. The possible gapped domain walls separating the two phases can be determined by employing the folding trick, see Ref.~\cite{Beigi2011}. This theorem states that the gapped domain walls between two arbitrary phases $D(G)$ and $D(G')$ are in one-to-one correspondence with the gapped boundaries of $D(G\times G')$, which can be parametrized by subgroups $K$ of $G\times G'$ and a 2-cocycle $\varphi\in H^2(K,\mathbb{C}^\times)$. In our case, there are 10 different subgroups of $\mathbb{Z}_2\times S_3$ up to conjugation, which are listed in Table~\ref{tab:subgroups}. Two of these subgroups additionally have a non-trivial 2-cocycle, which is why we count 12 different gapped domain walls between the $D(S_3)$ phase and the $D(\mathbb{Z}_2)$ phase in total. For simplicity, we restrict ourselves to considering the gapped domain walls corresponding to trivial 2-cocycles. Note that except for $K_4$ and $K_9$ all the subgroups in Table~\ref{tab:subgroups} can be written as a direct product of a subgroup of $\mathbb{Z}_2$ and a subgroup of $S_3$, which means that the corresponding gapped domain walls can simply be identified as a boundary (to the vacuum) for the $D(\mathbb{Z}_2)$ quantum double model placed next to a boundary for the $D(S_3)$ quantum double model. In order to determine the possible tunnellings of quasiparticles we consider the folded plane and determine which particles of the $D(\mathbb{Z}_2\times S_3)$ particle spectrum [which are effectively pairs consisting of a $D(\mathbb{Z}_2)$ and a $D(S_3)$ particle] condense to the vacuum at each boundary. Ref.~\cite{Beigi2011} provides an explicit algebraic expression to calculate these condensations. Table~\ref{tab:transitions} lists the possible tunnellings that were obtained in this way for each of the 10 different gapped domain walls corresponding to trivial 2-cocycles.

\begin{table}[bt]
	\centering
	\renewcommand{\arraystretch}{1.3}
	\begin{tabular}{|c||p{4cm}|c|c|}
		\hline
		&Subgroup&Order&Isomorphy class\\
		\hline\hline
		$K_1$&$\{e\}\times\{e\}$&$1$&trivial\\\hline
		$K_2$&$\{e\}\times\{e,t\}$&$2$&$\mathbb{Z}_2$\\\hline
		$K_3$&$\{e,x\}\times\{e\}$&$2$&$\mathbb{Z}_2$\\\hline
		$K_4$&$\{(e,e),(x,t)\}$&$2$&$\mathbb{Z}_2$\\\hline
		$K_5$&$\{e\}\times\{e,c,c^2\}$&$3$&$\mathbb{Z}_3$\\\hline
		$K_6$&$\{e,x\}\times\{e,t\}$&$4$&$\mathbb{Z}_2\times\mathbb{Z}_2$\\\hline
		$K_7$&$\{e,x\}\times\{e,c,c^2\}$&$6$&$\mathbb{Z}_6$\\\hline
		$K_8$&$\{e\}\times\{e,t,c,ct,c^2,c^2t\}$&$6$&$S_3$\\\hline
		$K_9$&$\{(e,e),(e,c),(e,c^2),(x,t),$\linebreak$(x,ct), (x,c^2t)\}$&$6$&$S_3$\\\hline
		$K_{10}$&$\{e,x\}\times\{e,t,c,ct,c^2,c^2t\}$&$12$&$\mathbb{Z}_2\times S_3$\\
		\hline
	\end{tabular}
	\caption{Subgroups of $\mathbb{Z}_2\times S_3$. Note that $K_4$ and $K_9$ cannot be written as a direct product of a subgroup of $\mathbb{Z}_2$ and a subgroup of $S_3$.}
	\label{tab:subgroups}
\end{table}

\begin{table}[b]
	\centering
	\renewcommand{\arraystretch}{1}
	\subfloat[][$K=K_1$]{
		\begin{tabular}{|c||c|c|c|c|c|c|c|c|}
			\hline
			&$A$&$B$&$C$&$D$&$E$&$F$&$G$&$H$\\
			\hline\hline
			$1$&1&1&2&&&&&\\
			\hline
			$e$&1&1&2&&&&&\\
			\hline
			$m$&&&&&&&&\\
			\hline
			$\epsilon$&&&&&&&&\\
			\hline
	\end{tabular}}
	\vspace{0.3cm}
	\hfill
	\subfloat[][$K=K_2$]{
		\begin{tabular}{|c||c|c|c|c|c|c|c|c|}
			\hline
			&$A$&$B$&$C$&$D$&$E$&$F$&$G$&$H$\\
			\hline\hline
			$1$&1&&1&1&&&&\\
			\hline
			$e$&1&&1&1&&&&\\
			\hline
			$m$&&&&&&&&\\
			\hline
			$\epsilon$&&&&&&&&\\
			\hline
	\end{tabular}}
	\hfill
	\subfloat[][$K=K_3$]{
		\begin{tabular}{|c||c|c|c|c|c|c|c|c|}
			\hline
			&$A$&$B$&$C$&$D$&$E$&$F$&$G$&$H$\\
			\hline\hline
			$1$&1&1&2&&&&&\\
			\hline
			$e$&&&&&&&&\\
			\hline
			$m$&1&1&2&&&&&\\
			\hline
			$\epsilon$&&&&&&&&\\
			\hline
		\end{tabular}
	}
	\vspace{0.3cm}
	\hfill
	\subfloat[][$K=K_4$]{
		\begin{tabular}{|c||c|c|c|c|c|c|c|c|}
			\hline
			&$A$&$B$&$C$&$D$&$E$&$F$&$G$&$H$\\
			\hline\hline
			$1$&1&&1&&&&&\\
			\hline
			$e$&&1&1&&&&&\\
			\hline
			$m$&&&&1&&&&\\
			\hline
			$\epsilon$&&&&&1&&&\\
			\hline
		\end{tabular}
	}
	\hfill
	\subfloat[][$K=K_5$]{
		\begin{tabular}{|c||c|c|c|c|c|c|c|c|}
			\hline
			&$A$&$B$&$C$&$D$&$E$&$F$&$G$&$H$\\
			\hline\hline
			$1$&1&1&&&&2&&\\
			\hline
			$e$&1&1&&&&2&&\\
			\hline
			$m$&&&&&&&&\\
			\hline
			$\epsilon$&&&&&&&&\\
			\hline
		\end{tabular}
	}
	\vspace{0.3cm}
	\hfill
	\subfloat[][$K=K_6$]{
		\begin{tabular}{|c||c|c|c|c|c|c|c|c|}
			\hline
			&$A$&$B$&$C$&$D$&$E$&$F$&$G$&$H$\\
			\hline\hline
			$1$&1&&1&1&&&&\\
			\hline
			$e$&&&&&&&&\\
			\hline
			$m$&1&&1&1&&&&\\
			\hline
			$\epsilon$&&&&&&&&\\
			\hline
		\end{tabular}
	}
	\hfill
	\subfloat[][$K=K_7$]{
		\begin{tabular}{|c||c|c|c|c|c|c|c|c|}
			\hline
			&$A$&$B$&$C$&$D$&$E$&$F$&$G$&$H$\\
			\hline\hline
			$1$&1&1&&&&2&&\\
			\hline
			$e$&&&&&&&&\\
			\hline
			$m$&1&1&&&&2&&\\
			\hline
			$\epsilon$&&&&&&&&\\
			\hline
		\end{tabular}
	}
	\vspace{0.3cm}
	\hfill
	\subfloat[][$K=K_8$]{
		\begin{tabular}{|c||c|c|c|c|c|c|c|c|}
			\hline
			&$A$&$B$&$C$&$D$&$E$&$F$&$G$&$H$\\
			\hline\hline
			$1$&1&&&1&&1&&\\
			\hline
			$e$&1&&&1&&1&&\\
			\hline
			$m$&&&&&&&&\\
			\hline
			$\epsilon$&&&&&&&&\\
			\hline
		\end{tabular}
	}
	\hfill
	\subfloat[][$K=K_9$]{
		\begin{tabular}{|c||c|c|c|c|c|c|c|c|}
			\hline
			&$A$&$B$&$C$&$D$&$E$&$F$&$G$&$H$\\
			\hline\hline
			$1$&1&&&&&1&&\\
			\hline
			$e$&&1&&&&1&&\\
			\hline
			$m$&&&&1&&&&\\
			\hline
			$\epsilon$&&&&&1&&&\\
			\hline
		\end{tabular}
	}\hspace{0.3cm}
	\vspace{0.3cm}
	\subfloat[][$K=K_{10}$]{
		\begin{tabular}{|c||c|c|c|c|c|c|c|c|}
			\hline
			&$A$&$B$&$C$&$D$&$E$&$F$&$G$&$H$\\
			\hline\hline
			$1$&1&&&1&&1&&\\
			\hline
			$e$&&&&&&&&\\
			\hline
			$m$&1&&&1&&1&&\\
			\hline
			$\epsilon$&&&&&&&&\\
			\hline
		\end{tabular}\hfill
	}
	\caption{The allowed tunnelling processes for the different gapped domain walls between the $D(\mathbb{Z}_2)$ phase and the $D(S_3)$ phase corresponding to subgroups of $\mathbb{Z}_2\times S_3$ and trivial 2-cocycles. A number in a given cell of the table indicates that the tunnelling of the $D(\mathbb{Z}_2)$ particle specified in the leftmost column to the $D(S_3)$ particle specified in the top row is possible (with the corresponding multiplicity given by that number), whereas an empty cell corresponds to the tunnelling multiplicity 0, i.e.~an impossible tunnelling. The labelling of the subgroups is defined in Table~\ref{tab:subgroups}.}
	\label{tab:transitions}
\end{table}

For brevity let us use the intuitive notation $x\rightleftarrows y$ if the tunnelling from an anyon $x$ of the phase $D(\mathbb{Z}_2)$ to an anyon $y$ of the phase $D(S_3)$ and vice versa is possible, and $x\centernot\rightleftarrows y$ if the tunnelling is impossible. In order to turn an $A+B+2C$ hole into a $1+e$ hole when it crosses the domain wall while mapping the logical state $|\psi\rangle=a|A\rangle+b|B\rangle$ to the corresponding state $|\psi\rangle=a|1\rangle+b|e\rangle$, we are interested in a domain wall where
\begin{align*}
1\quad\rightleftarrows\quad A,&&e\quad\rightleftarrows\quad B,
\end{align*}
but
\begin{align*}
1\quad&\centernot\rightleftarrows\quad B,&&e\quad\centernot\rightleftarrows\quad A,\\
1\quad&\centernot\rightleftarrows\quad C,&&e\quad\centernot\rightleftarrows\quad C.
\end{align*}
Looking at Table~\ref{tab:transitions}, we choose $K=K_9$ in order to realize our domain wall. The form of the corresponding domain wall stabilizers can easily be obtained from the general theory presented in~\cite{Beigi2011} and was given explicitly in Eqs.~(\ref{eq:boundaryvertexprojectorZ2S3}) and (\ref{eq:boundaryplaquetteprojectorZ2S3}).

As a possible simplification to our construction, recall that the plaquette operator $B^K_p$ acting on a single qubit and a single six-level spin explicitly reads
\begin{equation}
\begin{split}
B^K_p&=|e,e\rangle\langle e,e|+|e,c\rangle\langle e,c|+|e,c^2\rangle\langle e,c^2|\\&+|x,t\rangle\langle x,t|+|x,ct\rangle\langle x,ct|+|x,c^2t\rangle\langle x,c^2t|.
\end{split}
\end{equation} 
Therefore, if we focus on the case where we have no excitations on the domain wall plaquettes, it may be preferable to replace the two spins located at each edge along the domain wall by a single six-level spin and adjust the action of the operators that touch these spins appropriately. If we use the qubit-qutrit basis for the six-level systems in the $D(S_3)$ part of the lattice and also for those on the domain wall, these modified operators take on a particularly nice form. In fact, the modified plaquette operators $\tilde{B}_p^{\mathbb{Z}_2}$ along the $D(\mathbb{Z}_2)$ side of the boundary keep their form but now act on the qubit of the new six-level system instead of acting on the qubit that was originally placed on the domain wall. The modified plaquette operators $\tilde{B}^{S_3}_s$ along the $D(S_3)$ side of the boundary also keep their form, but act on the new six-level system instead of acting on the six-level system that was originally placed on the domain wall. The vertex operator $\tilde{A}^K_v$ now acts on three six-level systems, one in the $D(S_3)$ part of the lattice and two on the domain wall, as well as on a single qubit in the $D(\mathbb{Z}_2)$ part of the lattice. The corresponding stabilizer circuit can conveniently be realized using the qubit-qutrit construction of the six-level spins. In particular, let us place a six-level syndrome qudit at each domain wall vertex. We now realize the domain wall vertex circuits similar to Fig.~\ref{fig:circuits}(c), using the full vertex syndrome qudit as the control if a six-level spin is the target, and only its qubit part if the qubit is the target.

\section{Switching from the fusion space encoding to the hole encoding}

In order to inject the non-stabilizer state given in Eq.~(\ref{eq:nonstabilizer}) into the $D(\mathbb{Z}_2)$ phase, we have to switch from the fusion space encoding to a hole-pair encoding. In the following, let us point out some details regarding the fault-tolerance of this procedure. Throughout this section, we assume that non-Abelian error correction for the $D(S_3)$ model can be  performed on a sufficiently reliable level, even though an explicit proof of full fault-tolerance has still to be established. In order to distinguish between the fusion space encoding and the hole encoding, we will from now on use the subscript $f$ ($h$) for states corresponding to fusion channels (hole occupancies). Recall from Sec.~VI that we encode a qubit in (a subspace of) the fusion space of a quadruple of $G$ anyons with vacuum total fusion outcome by setting $|0\rangle_L=|A\rangle_f$ and $|1\rangle_L=|B\rangle_f$, where ($|0\rangle_L$, $|1\rangle_L$) denotes the computational basis. An arbitrary logical state in the fusion space encoding can then be written as $|\psi\rangle_f =a|A\rangle_f +b|B\rangle_f$. We want to transfer this state to the hole encoding given by the identification $|0\rangle_L =|A\rangle_h$, $|1\rangle_L=|B\rangle_h$ corresponding to the anyonic occupancy of two $A+B+2C$ holes.

\begin{figure*}[tb]
	\centering
	\includegraphics{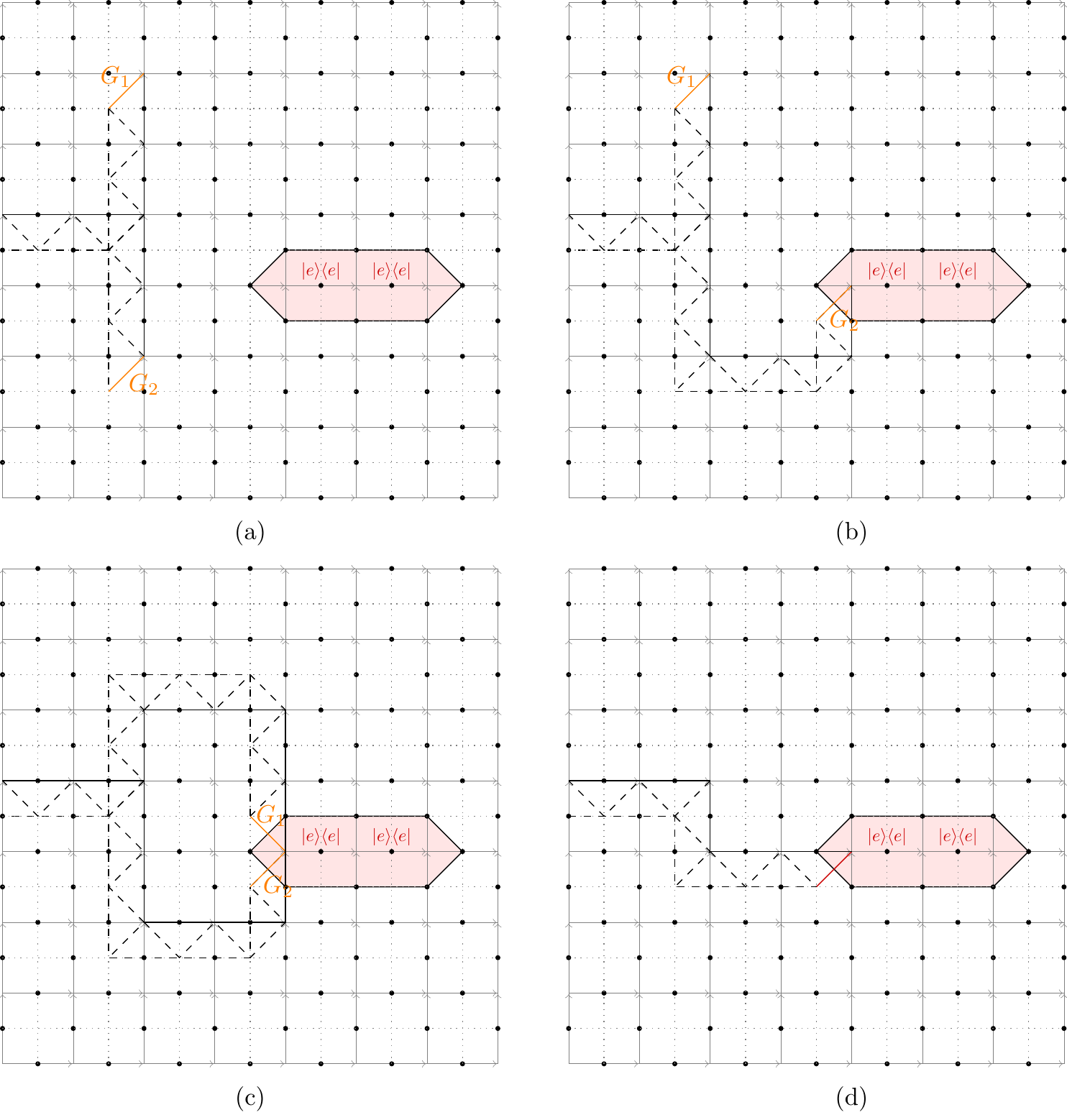}
	\caption{The $B$ occupancy of a pair of $G$ particles is transferred to an $A+B+2C$ hole. In the notations introduced in this Appendix, this maps the state $|\psi\rangle_f=a|A\rangle_f+b|B\rangle_f$ to the state $|\psi\rangle_h=a|A\rangle_h+b|B\rangle_h$. The dashed triangles indicate the support of the ribbon operators used to perform the movement of the $G$ particles. The shaded region denotes the hole, i.e. the region where the vertex stabilizers are turned off and replaced by single qudit measurements in the computational basis. (a) A pair of $G$ particles in the state $|\psi\rangle_f=a|A\rangle_f +b|B\rangle_f$ and an $A+B+2C$ hole with vacuum anyonic occupancy. (b) One of the $G$ particles is moved until its charge part is located inside the hole. (c) The second $G$ particle is moved into the hole as well and its charge part fused with the charge part of the first particle. (d) The flux parts of the two $G$ particles are fused, removing the $G$ particles from the lattice and leaving us with the hole in the state $|\psi\rangle_h=a|A\rangle_h +b|B\rangle_h$.}
	\label{fig:switch}
\end{figure*}

Consider the situation that is shown in Fig.~\ref{fig:switch}(a), where we have a pair of $G$ particles in an arbitrary state $|\psi\rangle_f =a|A\rangle_f +b|B\rangle_f$ and an empty $A+B+2C$ hole. Since the hole encoding (the fusion space encoding) is topologically protected in the case of large holes (well-separated particles), the only critical steps in the transfer of information include the fusion of the $G$ particles with the holes. Throughout this section we assume that the two pairs of $G$ particles (the two $A+B+2C$ holes) are located far away from each other, which means that tunnellings of quasiparticles from one pair to the other (one hole to the other) are sufficiently unlikely.  We are thus only concerned with logical $Z$ errors. Let us now move one of the $G$ particles towards the hole until its charge part lies within the boundary of the hole, see Fig.~\ref{fig:switch}(b). It is instructive to explicitly consider the logical $Z$ operator (\ref{eq:logZ}), which now has to enclose the pair of $G$ particles as well as the hole in order not to leave a local trace on the disentangled qudits within the hole. As such, the $A$ and $B$ particles are now delocalized between both the pair of $G$ particles and the vertices that are part of the hole. This situation becomes even clearer when we assume that the qudits inside the hole are removed from the code (with the plaquette stabilizers along the boundary of the hole modified accordingly). In this picture it is evident that the charge part of the $G$ particle is delocalized among all vertices comprising the hole as soon as it enters the hole. We can now do the same for the other $G$ particle and move its charge part into the hole as well, effectively fusing it with the charge part of the first $G$ particle. This situation is shown in Fig.~\ref{fig:switch}(c). Finally we annihilate the two $G$ particles by fusing their flux parts, see Fig.~\ref{fig:switch}(d). Note that the operations needed to move the $G$ particles act locally on this pair of particles and in particular do not have support on a ribbon enclosing the hole. Repeating the same steps for the second pair of $G$ particles and a second empty $A+B+2C$ hole leaves us with two $A+B+2C$ holes which encode the logical state $|\psi\rangle_h =a|A\rangle_h +b|B\rangle_h$. Furthermore, one can show that the projection onto the $G$ fusion channel of the pair of $G$ particles commutes with the single qudit terms acting inside the hole, as a $G$ particle is not delocalized among the vertices comprising the hole. Similarly, any $C$ particles that might accidentally enter the hole do not get delocalized among the $G$ particles.

It would now be interesting to consider the reverse process, i.e.~switching from the hole encoding to the fusion space encoding. However, this is not as easy as it might seem at first sight, as the fusion space encoding is inherently stronger than the hole encoding. While the former is only accessible by truly non-local operations, the logical state in the hole encoding is accessible by a local operations and classical communication (LOCC) protocol that simply measures the anyonic occupancy of all the vertices comprising the hole independently and then takes the parity of the number of measured $B$ anyons. The obvious analogue to the process described in Fig.~\ref{fig:switch} would be to create a pair of $G$ anyons from the vacuum, fuse one of these particles with the hole that is in the logical state $|\psi\rangle_h=a|A\rangle_h+b|B\rangle_h$ and then contract the hole in order to completely transfer its anyonic occupancy to the pair of $G$ particles. However, there is one major drawback to this procedure: Fusing a $B$ particle with a $G$ particle will in general cause a rotation in the local subspace carried by that $G$ particle. Only the braiding operation leaves the local degrees of freedom invariant as long as the particles are well separated at all times; fusion with an additional particle, even though performing a rotation in the topologically protected space as well, may have an unwanted side effect that is locally detectable. In order to make this explicit, let us consider two different bases for the local subspace carried by a $G$ particle that is located at a site $s=(p,v)$. One possible choice is given by $|c\rangle_{G,l}, |c^2\rangle_{G,l}$, where we use the subscripts $G$ to denote the particle type and $l$ to indicate that we are dealing with local degrees of freedom. Projectors onto these local basis states are given by $B^c_{s}P_s^G$ and $B^{c^2}_{s}P_s^G$, see Eqs.~(\ref{eq:plaquetteoperator}) and (\ref{eq:Gprojector}). Fusing a $B$ particle with the $G$ anyon involves applying the operator $F^B_j$, see Eq.~(\ref{eq:Bribbonop}), to a single spin $j$ located at an edge connecting the vertex $v$ to an adjacent one. Obviously, this operation commutes with the projectors onto the local states. However, let us consider a different set of basis states for the internal degrees of freedom, namely the basis given by the states $|c\rangle_{G,l}+|c^2\rangle_{G,l}, |c\rangle_{G,l}-|c^2\rangle_{G,l}$. The corresponding projectors are given by $(B^c_s+B^{c^2}_s+A^t_vB^c_s+A^t_vB^{c^2}_s)P^G_s$, $(B^c_s+B^{c^2}_s-A^{t}_vB^c_s-A^t_vB^{c^2}_s)P^G_s$ up to a constant factor. In this case, the single spin operation $F^B_j$ flips the basis states as it anticommutes with the $A^t_v$ terms. As such, the fusion with a $B$ anyon leaves a trace in the local subspace of the corresponding $G$ particle, which makes the information stored in the hole locally accessible when the hole is contracted. An approach that might be able to overcome this difficulty could be to use more than one pair of $G$ particles. In particular, one could imagine to transfer the anyonic occupancy of each vertex of the hole to a separate pair of $G$ particles. By preparing all the $G$ particles in a flux state of either $|c\rangle_{G,l}$ or $|c^2\rangle_{G,l}$ and assuming that local perturbations occur only with low probability, it might be possible to distil a single pair of $G$ particles such that the anyonic occupancy of the hole is encoded non-locally in the fusion space of this single pair. We leave such considerations to future work. Note that if fusion with a $B$ particle is used to perform a logical $X$ operation on the qubit encoded in a quadruple of $G$ anyons of known local state, the unwanted local rotation can in principle be reverted by applying a suitable correction operator after every fusion. This was contemplated in Ref.~\cite{Wootton2009} in the case of $C$ anyons, which show a similar behaviour upon fusion with $B$ anyons.

\section{Transporting logical information across the domain wall}

Let us now comment upon some details regarding the fault-tolerance of the transport of logical information across the domain wall. Again, we assume that there is a way to perform non-Abelian error correction for the $D(S_3)$ model in a sufficiently reliable way. Given that large holes enjoy an intrinsic topological protection, the only critical steps in the transfer are those shown in Figs.~3(a) and 3(c). Again we assume that the second hole used to encode the logical qubit is situated sufficiently far away, such that we are only concerned with logical $Z$ errors. Let us start by discussing the process shown in Fig.~3(a). By construction of the domain wall, the encoded information remains delocalized among \emph{all} vertices of the hole, including the domain wall vertex $v_d$, when the hole is extended. It is instructive to verify this explicitly by noting that the logical $Z$ operator [see Eq.~(\ref{eq:logZ})] at this stage of the transfer has to act on both the $D(S_3)$ part of the hole as well as on $v_l$ in order not to leave a local trace on the domain wall. The situation becomes even clearer when we use the reduced domain wall construction described in Appendix C, where the domain wall plaquettes are fixed in their ground state and implemented by six-level systems instead of a combination of a qubit and a six-level system. As such it is not possible to act on $v_l$ and $v_r$ separately from a hardware point of view. In either case, an operator causing a logical $Z$ error has to enclose both parts of the domain wall vertex $v_d$. Explicitly, the logical state of the hole is now given by the net $B$ occupancy of the $D(S_3)$ part as well as the net occupancy of the vertex $v_d$ in terms of pairs $(1,B)$ and $(e,A)$, which are indistinguishable. Let us now turn to the process shown in Fig.~3(c), where we measure the stabilizer corresponding to the domain wall vertex $v_d$ in order to fully contract the $D(S_3)$ part of the hole. An outcome corresponding to a non-trivial quasiparticle is corrected by moving the quasiparticle into the hole. Particular interest should be paid to the fact that this measurement can yield an outcome corresponding to a pair $(1,C)$ or $(e,C)$ located on $v_d$. Let us emphasize that these two pairs are indistinguishable. Since the $C$ anyons cannot enter the $D(\mathbb{Z}_2)$ phase it might not be clear how to correct for such a measurement outcome. However, recall that we are only interested in moving $A+B+2C$ holes across the boundary whose anyonic occupancy is given by a state $|\psi\rangle_h\in \mathrm{span}(|A\rangle_h,|B\rangle_h)$. This means that any $C$ particle that might reside on the boundary vertex $v_d$ after the final contraction process must have entered the hole as the consequence of an error in the $D(S_3)$ part of the hybrid model, and another stray $C$ anyon (or a collection of stray anyons with total fusion channel $C$) has to be located somewhere in the $D(S_3)$ phase. Since we assume the probability for such errors to be small and we correct for errors on a regular basis, this $C$ anyon will be located close to the vertex $v_d$. It is now crucial to fuse this anyon with the one residing at $v_d$ in order to recover any hidden $B$/$e$ anyon, which must then be fused with the hole to preserve its logical state.

The reverse process can be realized by reverting the steps (a) to (c) shown in Fig.~3 and arguing in a similar way as above.


\section{Completing the gate set}
\label{subsec:universalTQC}

This Appendix is a short summary of a particular result presented in Ref.~\cite{Bravyi2005}. With the tools that were developed in the main text it is possible to prepare the state $|\psi\rangle_f=\cos\left(\frac{2\pi}{3}\right)|A\rangle_f+i\,\sin\left(\frac{2\pi}{3}\right)|B\rangle_f$ in the $D(S_3)$ phase, transform this state into $|\psi\rangle_h=\cos\left(\frac{2\pi}{3}\right)|A\rangle_h+i\,\sin\left(\frac{2\pi}{3}\right)|B\rangle_h$ and finally inject the state into the $D(\mathbb{Z}_2)$ phase, where it emerges as
\begin{equation}
|\psi\rangle=\cos\left(\frac{2\pi}{3}\right)|1\rangle+i\,\sin\left(\frac{2\pi}{3}\right)|e\rangle.
\end{equation}
In particular, we identify the logical basis in the $D(\mathbb{Z}_2)$ phase as $|0\rangle_L=|1\rangle$, $|1\rangle_L=|e\rangle$. In the following, however, we are going to drop the subscript $L$ for brevity. Let us now briefly summarize how many copies of the state $|\psi\rangle$ can be used to perform the non-Clifford unitary
\begin{equation}
U=\begin{pmatrix}
1&0\\
0&e^{\frac{2\pi i}{3}}
\end{pmatrix}
\end{equation}
on a logical qubit given that we are able to fault-tolerantly perform the following operations: i) initialize logical qubits in the state $|0\rangle$, ii) perform arbitrary Clifford operations on logical qubits and iii) measure logical qubits in any Pauli basis. All of these assumptions are justified in a standard implementation of the $D(\mathbb{Z}_2)$ surface code using a $1+e$ hole-pair encoding. From now on, let us refer to the qubit corresponding to the state $|\psi\rangle$ as the ancillary qubit. The unitary operation $U$ can then be performed on a qubit in the arbitrary state $|\varphi\rangle=a|0\rangle+b|1\rangle$, which we are going to call the computational qubit, by the following steps:

\begin{enumerate}
	\item Apply a Hadamard gate $H=\frac{1}{\sqrt 2}\begin{pmatrix}1&1\\
	1&-1\end{pmatrix}$ to the state $|\psi\rangle$ in order to obtain the state
	\begin{equation}
	|\psi'\rangle=H|\psi\rangle=\frac{e^{\frac{2\pi i}{3}}}{\sqrt{2}}\left(|0\rangle+e^{\frac{2\pi i}{3}}|1\rangle\right).
	\end{equation}
	By ignoring the global phase this is equivalent to the state
	\begin{equation}
	|\psi'\rangle=\frac{1}{\sqrt{2}}\left(|0\rangle+e^{\frac{2\pi i}{3}}|1\rangle\right).
	\end{equation}
	\item Prepare the state $|\Psi_0\rangle=|\varphi\rangle\otimes|\psi'\rangle$.
	\item Measure $\sigma^z\otimes\sigma^z$. The possible measurement outcomes are $+1$ and $-1$, both occurring with probability $1/2$. If the outcome is $+1$, we are left with the state
	\begin{equation}
	|\Psi_1^+\rangle=a|00\rangle+be^{\frac{2\pi i}{3}}|11\rangle.
	\end{equation}
	If the measurement outcome is $-1$, the resulting state is
	\begin{equation}
	|\Psi_1^-\rangle=ae^{\frac{2\pi i}{3}}|01\rangle+b|10\rangle.
	\end{equation}
	\item Apply a $\mathrm{CNOT}$ gate with the computational qubit as the control and the ancillary qubit as the target. The resulting states are
	\begin{align}
	|\Psi_2^+\rangle&=\left(a|0\rangle+be^{\frac{2\pi i}{3}}|1\rangle\right)\otimes|0\rangle,\\
	|\Psi_2^-\rangle&=\left(ae^{\frac{2\pi i}{3}}|0\rangle+b|1\rangle\right)\otimes|1\rangle,
	\end{align}
	which leaves the ancillary qubit disentangled from the system.
	\item Discard the ancillary qubit. The net effect of this procedure is thus the application of either $U$ or $U'$ to the computational qubit, with
	\begin{equation}
	U=\begin{pmatrix}
	1&0\\
	0&e^{\frac{2\pi i}{3}}
	\end{pmatrix},\qquad
	U'=\begin{pmatrix}
	1&0\\
	0&e^{-\frac{2\pi i}{3}}
	\end{pmatrix}.
	\end{equation}
	\item By repeating this procedure for many copies of the ancillary state $|\psi\rangle$ we will eventually succeed in realizing the gate $U$, and the probability that we require more than $N$ steps in order to achieve this decreases exponentially with $N$. For more details on this argument, the reader is referred to the original publication~\cite{Bravyi2005}.
\end{enumerate}
Finally, it is shown in Ref.~\cite{Nebe2001} that the full Clifford group supplemented with a single arbitrary non-Clifford gate allows for universal quantum computation.

\bibliographystyle{unsrt}

\end{document}